\documentclass[scr,twocolumn,superscriptaddress,floatfix]{revtex4-1}

\usepackage{graphicx}
\usepackage{tabularx}
\usepackage{subfigure}
\usepackage{amssymb}
\usepackage{enumerate}
\usepackage{array}

\usepackage[normalem]{ulem} 
\newlength{\gnupicwidth}
\usepackage{color}

\definecolor{rot}{rgb}{1,0,0}
\definecolor{blau}{rgb}{0,0,1}
\definecolor{orange}{rgb}{.5,.5,0}
\definecolor{dunkelgruen}{rgb}{.133,0.545,0.133}

\begin{document}
\title{A  versatile UHV transport and measurement chamber for neutron reflectometry under UHV conditions}

\author{A. Syed Mohd}
\email[Electronic mail:]{ A.Syed.Mohd@fz-juelich.de}
\affiliation{J{\"u}lich Centre for Neutron Science (JCNS)
at Heinz Maier-Leibnitz Zentrum (MLZ)
Forschungszentrum J{\"u}lich GmbH,
Lichtenbergstr.\ 1,
85747 Garching, Germany
}

\author{S. P{\"u}tter}
\email[Electronic mail:]{S.Puetter@fz-juelich.de}
\affiliation{J{\"u}lich Centre for Neutron Science (JCNS)
at Heinz Maier-Leibnitz Zentrum (MLZ)
Forschungszentrum J{\"u}lich GmbH,
Lichtenbergstr.\ 1,
85747 Garching, Germany
}

\author{S. Mattauch}
\affiliation{J{\"u}lich Centre for Neutron Science (JCNS)
at Heinz Maier-Leibnitz Zentrum (MLZ)
Forschungszentrum J{\"u}lich GmbH,
Lichtenbergstr.\ 1,
85747 Garching, Germany
}
\author{A. Koutsioubas}
\affiliation{J{\"u}lich Centre for Neutron Science (JCNS)
at Heinz Maier-Leibnitz Zentrum (MLZ)
Forschungszentrum J{\"u}lich GmbH,
Lichtenbergstr.\ 1,
85747 Garching, Germany
}
\author{H. Schneider}
\affiliation{J{\"u}lich Centre for Neutron Science (JCNS)
at Heinz Maier-Leibnitz Zentrum (MLZ)
Forschungszentrum J{\"u}lich GmbH,
Lichtenbergstr.\ 1,
85747 Garching, Germany
}
\author{A. Weber}
\affiliation{J{\"u}lich Centre for Neutron Science (JCNS)
at Heinz Maier-Leibnitz Zentrum (MLZ)
Forschungszentrum J{\"u}lich GmbH,
Lichtenbergstr.\ 1,
85747 Garching, Germany
}
\author{T. Br{\"u}ckel}
\affiliation{J{\"u}lich Centre for Neutron Science (JCNS)
at Heinz Maier-Leibnitz Zentrum (MLZ)
Forschungszentrum J{\"u}lich GmbH,
Lichtenbergstr.\ 1,
85747 Garching, Germany
}
\affiliation{J{\"u}lich Centre for Neutron Science (JCNS) and Peter Gr{\"u}nberg Institute (PGI); JCNS-2, PGI-4: Scattering Methods
Forschungszentrum J{\"u}lich GmbH,
52425 J{\"u}lich, Germany
}




\date{December 29, 2016}

\begin{abstract} 
We report on a versatile mini ultra high vacuum (UHV) chamber which is designed
 to be used on  the MAgnetic Reflectometer with high Incident Angle (MARIA) of the J{\"u}lich Centre for Neutron Science at Heinz Maier-Leibnitz Zentrum in Garching, Germany. Samples are prepared in the  adjacent thin film laboratory by molecular beam epitaxy and moved into the compact  chamber for transfer without exposure to ambient air.
The  chamber  is based on DN 40 CF flanges and  equipped with sapphire view ports, a small getter pump and a 
wobble stick, which serves also as  sample holder.  

Here, we present polarized neutron reflectivity measurements which have been performed on Co thin films  at room temperature in UHV and in ambient air in a magnetic field of 200 mT and in the Q-range of 0.18 \AA$^{-1}$. The results confirm that
 the Co film is not contaminated during the polarized neutron reflectivity measurement. Herewith it is demonstrated that the mini UHV transport chamber also works  as measurement chamber  which opens new possibilities for  polarized neutron measurements under UHV conditions.

\end{abstract}


\maketitle

\section{Introduction}
The investigation of thin film samples which are sensitive to ambient
air with instruments at large scale facilities is challenging.
While simple sputter chambers for
 thin film growth and
\emph{in-situ} measurements have been realized, complex and much more powerful ultra high vacuum (UHV) systems, e.g.\ for molecular beam epitaxy (MBE) with several elements  cannot be placed on-site due to limited space because the movement of the detector has to be taken into account. Further restrictions appear for neutron reflectivity measurements as a magnet for the magnetic field is required and the sample has to be  in the UHV chamber as well as in the center of the yoke of the magnet.

Only  few growth chambers are reported for neutron experiments.\cite{Nawr99,Dura09,Kreu14} As an alternative approach, 
ultra--thin films which are sensitive to ambient air are subsequently
covered  by protective cap layers  after fabrication. However, this
 may change the physical properties of the thin film.
\cite{Call05,Hutt08}  Another solution is the
installation of a small UHV chamber only dedicated to measurements at the
instrument.\cite{Slob12} Consequently, a  UHV transfer chamber is required for bridging the gap
between the  UHV preparation system and the measurement chamber.\cite{Jiri98,Firp05,Renn07,Wata16}  However, this method requires two transfers
from one UHV chamber to the other which is inconvenient and time consuming.

 We  developed a mini
UHV chamber made from standard UHV parts which solves these problems. The chamber is capable of both, sample transfer and neutron reflectivity measurement under UHV conditions. 
An excellent candidate for testing the potential of the UHV chamber  is the neutron
reflectometer MARIA   at MLZ in Garching, Germany  designed for the investigation of ultra--thin magnetic films.\cite{Matt15} 

 In this paper, we show the versatility 
of this chamber. As test sample  a 30~\AA Co film
  on Pt(205 \AA)/MgO(001)  was produced in the MBE system located in  an adjacent building to the neutron reflectometer's location. Co was chosen due to its high reactivity with air. We present the first  reflectometry measurements under UHV conditions at MARIA. The results reveal also the feasibility of the mini UHV transport chamber  as a measurement chamber for neutron reflectometry.

\section{Experimental}

UHV substrate preparation and thin film growth were performed in a DCA
M600 MBE system with a base pressure of $10^{-10}$~mbar.  A $1 \times
1$ cm$^2$ MgO(001) single crystal (Crystec GmbH, Berlin, Germany)
served as substrate. To get rid of the H$_2$O at the surface, the MgO(001) was annealed at 200$^\circ$C  under UHV conditions for 2 hours. Subsequently,  a 205 \AA\  Pt buffer layer was deposited at 70$^\circ$C followed by growth of 30~\AA\ Co  at 50$^\circ$C. The Pt buffer  prevents the Co film from oxidation and intermixing  with MgO.\cite{Lalo08} The Pt was grown using e-beam evaporation  at a rate of  3 \AA/min.  The deposition rate of Co  was kept at 1.2 \AA/min using a high temperature effusion cell. During growth the pressure did not exceed
$4 \cdot 10^{-10}$ mbar.
The evaporation rate was monitored using a precalibrated quartz microbalance.

The MAgnetic Reflectometer with high Incident Angle (MARIA) of the J{\"u}lich Centre for Neutron Science
at Heinz Maier-Leibnitz Zentrum  
in Garching, Germany, has a horizontal scattering plane. It provides polarized neutron reflectometry (PNR) in standard operation. 
At the sample position, a hexapod with a Bruker electromagnet 
is installed. In-plane magnetic fields of maximum 1.2 T can be applied. For the transfer chamber  the pole shoes have to be unmounted which  reduces the magnetic field to max.\ 600~mT at a gap of about 12~cm. The measurements are performed at room temperature.
For  data treatment of
 the PNR  measurements  the program GenX is used.\cite{Bjoe07}
\section{System layout}
\label{sec:sys-conf}

\begin{figure}
\includegraphics[width=\gnupicwidth]{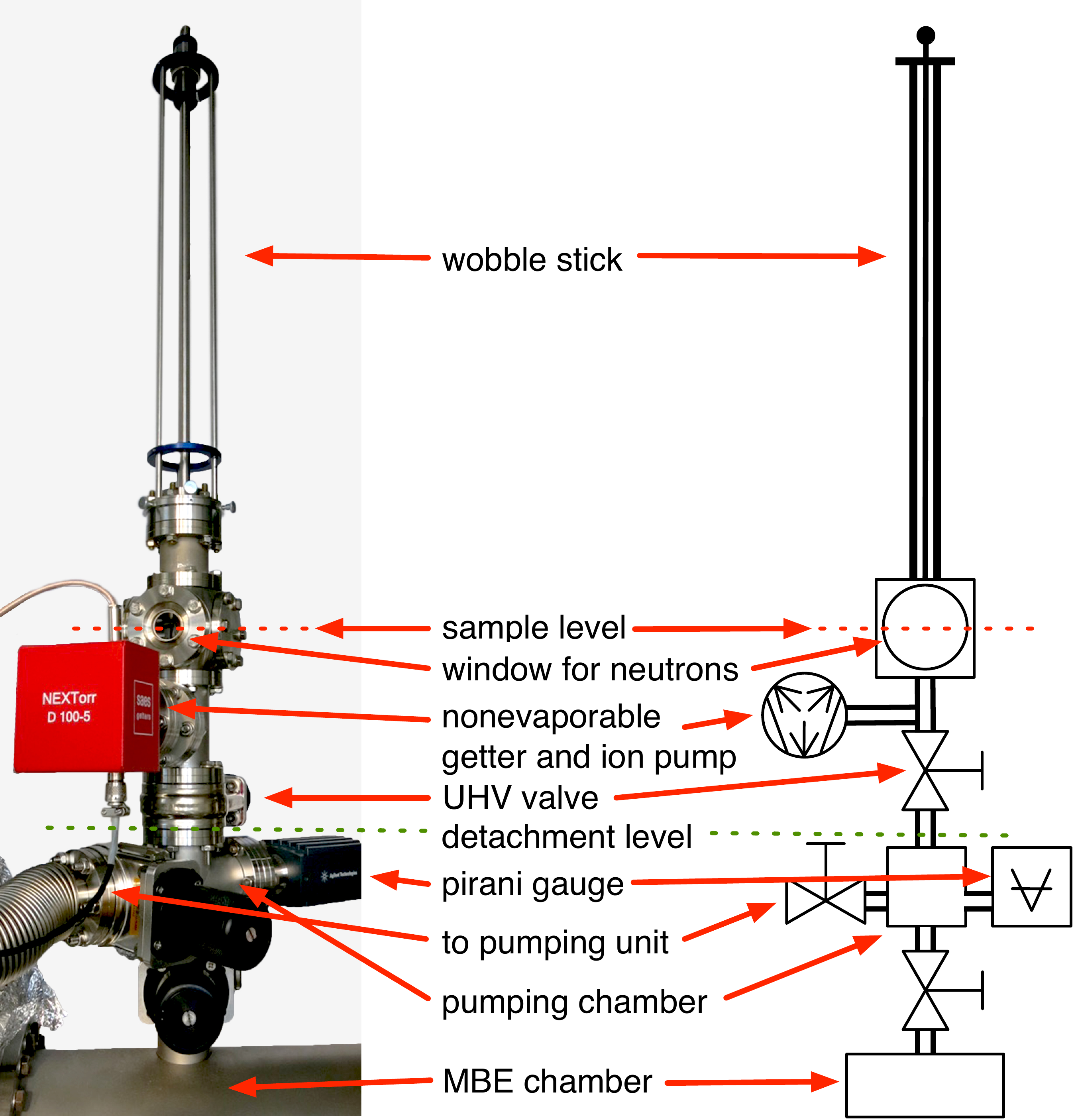}
 \caption{\label{fig:setup}(Color online) Photo (left) and sketch (right) of the transfer chamber attached to the MBE setup.}
 \end{figure}

A photograph and a sketch of the transfer chamber are shown in Fig.~\ref{fig:setup}. 
 The transportable chamber is lightweight (about 10 kg) and easy to transport.  Special
 attention has been paid to the compatibility of the chamber to
 neutron reflectivity measurements. Mostly standard components with DN
 40 CF (CF ConFlat, a registered trademark of Varian, Inc.) have been
 chosen in preferably non-ferromagnetic version.  The core part is a
 DN 40 CF cube made of low magnetic permeability  steal (316LN). Two opposing DN 40 CF sapphire
 view ports are mounted for enabling a clear neutron beam path. The maximum angle of incidence is 13$^\circ$ for the neutron beam which is  sufficient for measurements up to $Q_z=0.3$ \AA$^{-1}$ at $\lambda =4.5$~\AA. Here $Q_z$ denotes the component of the scattering vector perpendicular to the sample surface. To reach UHV the transport chamber is baked at 150$^\circ$C for 48 hours. 

\begin{figure}
\includegraphics[width=.5\gnupicwidth]{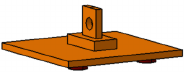}
 \caption{\label{fig:sample-plate}(Color online) Sketch of the sample  holder ($2 \times 2$ cm$^2$) with small handle on the back. The sample is attached facing downwards. }
 \end{figure}

For easy handling, the sample is attached to a sample holder which is sketched in Fig.~\ref{fig:sample-plate}. On its front side the sample is clamped with bolt heads. On the back side the sample holder has  a small handle for being grabbed. The typical sample size is $1 \times 1$ cm$^2$. Our  sample holder is made from Cu. However, any UHV compatible material is possible.

 For inserting the sample holder with the attached sample from
 the MBE setup into the transfer chamber a wobble stick (Ferrovac
 GmbH) is used. In general, a wobble stick serves for seizing
   things and translating and rotating them inside a UHV
   chamber. Here, together with the sample holder, it also serves for
   keeping the sample in position for measurement after retracting it
   into the transfer chamber. The wobble stick  can be
 fixed mechanically from outside to prevent any sample movement. For
 the given orientation of the transfer chamber in Fig.~\ref{fig:setup}
 the sample is facing downwards.

\begin{figure}
\includegraphics[width=\gnupicwidth]{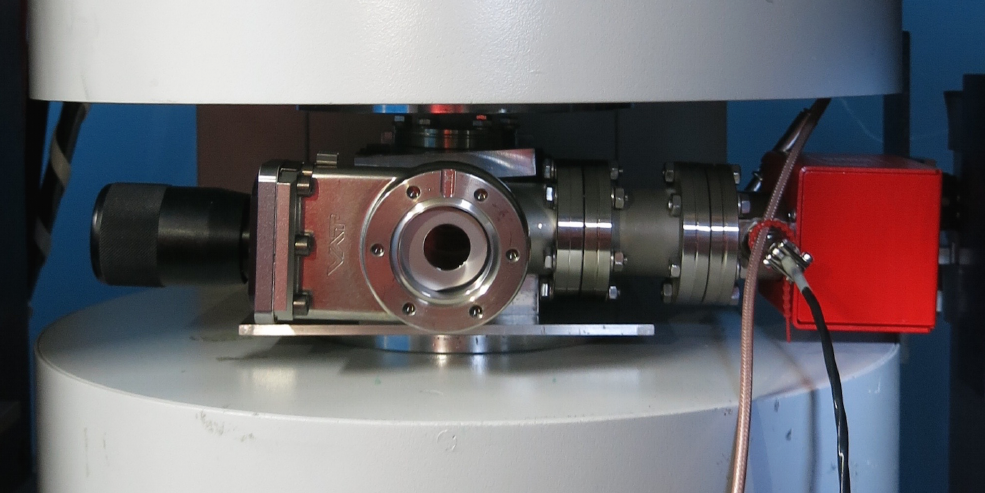}
 \caption{\label{fig:setup-maria}(Color online) Photo of the transfer chamber inside the MARIA magnet. In the center, the valve with the customized window is visible. On the right the ion getter pump is located.}
 \end{figure}

 Opposite to the wobble stick a DN 40 CF UHV tee is
 attached. At 90 degrees from it, a getter pump type Nextorr D 100-5
 (SAES Getters SpA, Italy) is installed. The pump is a combination of
 a sputter ion pump (SIP) and NEG material (nonevaporable getter material)
 \cite{Park11}  with a maximum pumping speed  of 140 l/s.  The pump is encased by a $\mu$-metal shield to
 minimize the magnetic stray field, which is about 0.08 mT at the sample position. 
 This value is well below the remanence of
 the magnet of the neutron reflectometer and the guide field of 0.2 mT. Therefore it does not influence the measurements. Two power supplies are used for ion pump operation, one is placed
 at the MBE system and the other one is located at the neutron
 reflectometer. During transfer, only the NEG element is pumping. If necessary, permanent operation of the ion pump can be supplied. The
 pressure is measured by the ion getter pump. The base pressure of the transfer chamber is
 about $2 \cdot 10^{-10}$~mbar.  The chamber is closed by a customized
 DN 40 CF valve (VAT GmbH) which has a small quartz window inside. This
 window allows for prealignment of the sample inside transfer chamber thereby observing a reflected laser beam, Fig.~\ref{fig:setup-maria}.

\section{Procedure of sample transfer}

\subsection{Attachment to the MBE setup}
\label{sec:attach-mbe}
At the MBE setup, the transfer chamber is attached to a  DN 40 CF cross with a valve, a Pirani vacuum gauge and a pumping unit, Fig.~\ref{fig:setup}. 
Within the MBE  setup samples are moved horizontally in a trolley, facing
downwards.  For the sample transfer a  trolley is moved below the transfer chamber.
  A window in the
 UHV chamber at the level of the trolley enables the control of the
transfer. After opening the valves the wobble stick of the transfer chamber is moved downwards to the trolley. The pincer grip of the wobble stick grabs the handle of the sample
 (Fig.~\ref{fig:sample-plate}) and by withdrawing the wobble stick into the transfer chamber
the sample is moved to its  position. After
closing the valve and venting the UHV cross the transfer chamber can
be detached from the MBE system and carried to the neutron
instrument.

\subsection{Neutron reflectivity measurements at MARIA}
\label{sec:Meas-maria}

The transfer chamber is transferred to the neutron instrument within approximately 5 minutes.
It is inserted horizontally between the pole shoes of the magnet, Fig.~\ref{fig:setup-maria}, and the pump is attached to its power supply. The pressure does not change during this procedure. After keeping the transfer chamber without
power for 24 hours the pressure rises  up to $1.3\cdot 10^{-9}$ mbar. For sensitive samples constant pumping is enabled.
 The PNR measurements
 are performed at room temperature.
A typical measurement of spin up (R+) and spin down (R-) reflectivity usually takes three hours.

\section{Results and discussion}
\label{sec:results}

\subsection{PNR-Measurements}
\label{sec:PNR}

\begin{figure}
\includegraphics[width=\gnupicwidth]{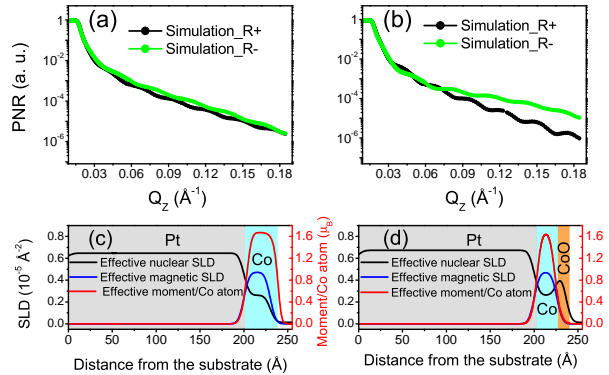}
 \caption{\label{fig:PNR-Co}(Color online) Simulated PNR curves of (a) Co film:  30~\AA~Co/205~\AA~Pt/MgO(001)  and (b) a partially oxidized Co film:  10~\AA~CoO/25~\AA~Co/205 \AA~Pt/MgO(001). In (c) and (d)  the nuclear and magnetic SLD profiles of the reflectivity curves in (a) and  (b) as well as the effective moment per Co atom are shown. }
 \end{figure}
In this part, we present polarized neutron reflectivity (PNR) measurements  of a 30~\AA\ Co film which was first prepared in the MBE setup and then measured in the transport chamber under UHV conditions at MARIA. Afterwards the same sample was measured again in ambient conditions after taking it out of the transfer chamber.

For comparison, we first discuss simulated
  PNR curves of 30~\AA~Co/205~\AA~Pt/MgO(001)   and  10~\AA~CoO/25~\AA~Co/205 \AA~Pt/MgO(001) in Figs.~\ref{fig:PNR-Co}(a) and (b), respectively. 
 While the nuclear Scattering Length Density (nuclear SLD) is always the same the
  magnetic Scattering Length Density (magnetic SLD) depends on the in-plane
  magnetization of the sample  parallel  or anti-parallel to  the applied field. As a
  result separate  spin up (R+) and the spin down (R--)
  channels give  information about the magnetic profile of
  the film.  
In the simulation with GenX \cite{Bjoe07} we utilized the nominal nuclear and magnetic scattering
  length density of MgO, Pt, Co and CoO, respectively. The film roughness  was set to  a typical value of 5 \AA\ at each interface.

 The resulting  effective nuclear and magnetic SLD depth profiles as well as the effective magnetic moment per Co atom are shown in Figs.~\ref{fig:PNR-Co}(c) and (d), respectively. In PNR, roughness at the interface is reflected as  a continuous variation of the SLD across the interface. As a consequence, the SLDs and the effective magnetic moment exceed the nominal film thicknesses  given  by the coloured background  as guide to the eye in the figures.

For the profile of the effective magnetic moment per Co atom, we have normalized the magnetic SLD with the nuclear SLD utilizing the formula of Ankner et 
al.~\cite{Ankn93b}:
\begin{equation}
 \frac{ \mathrm{Magnetic\, moment}}{\mathrm{atom}}=\frac{b}{c}\cdot \frac{\mathrm{Magnetic\, SLD}}{\mathrm{Nuclear\, SLD}} \nonumber
\end{equation}
where $b$ is the scattering length of the atom and $c=2.69\cdot 10^{-5}\mathrm{~\AA}/\mu_\mathrm{B}$. The simulation in Fig.~\ref{fig:PNR-Co} clearly  shows that the PNR measurement of an oxidized Co film will differ from that of a pure Co film.

\begin{figure}[t]
\includegraphics[width=\gnupicwidth]{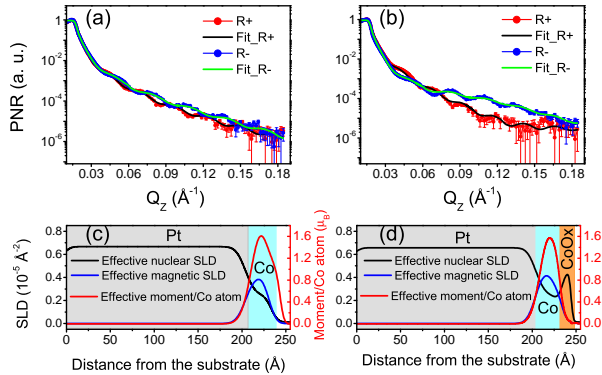}
 \caption{\label{fig:PNR-CoO}(Color online) PNR curves of the 30~\AA\ Co film (a) measured under UHV conditions utilizing the UHV transport chamber and (b) measured after exposing the same Co film to ambient air for 6 hours. In (c)  and (d) the neutron SLD  profiles obtained from the fitting of the PNR data measured under UHV conditions and in ambient air are shown.}
 \end{figure}

In Fig.~\ref{fig:PNR-CoO} we present PNR data of  Co/Pt/MgO(001)  measured  first in UHV utilizing the transport chamber and second  \textit{ex-situ} after exposing it to ambient air for about  6 hours, see Fig.~\ref{fig:PNR-CoO}(a) and (b), respectively.  
The pressure in the transport chamber during the measurement in UHV was $4 \cdot  10^{-10}$~mbar. The measurements were performed at room temperature with a magnetic field of 200~mT, which is well above the saturation field of Co thin films of  10~mT~\cite{Blan91,Puet01}.
The direction of the magnetic field was kept parallel to the polarisation axis of the neutron beam. Already a qualitative comparison of  Fig.~\ref{fig:PNR-CoO}(a) and (b)  reveals differences in the reflectivity curves which are  in good accordance with the ones in the simulated curves, Fig.~\ref{fig:PNR-Co}. 

%
\newcolumntype{L}[1]{>{\raggedleft\let\newline\\\arraybackslash\hspace{0pt}}m{#1}}
\begin{table}[t]
\begin{center}
\setlength{\extrarowheight}{2pt}
\begin{tabular}{
|m{.17\linewidth}*{4}{|>{\raggedleft}m{.17\linewidth}}p{.00001\linewidth}|}
\hline
& \multicolumn{2}{m{.27\linewidth}|}{in UHV\newline ($4 \cdot 10^{-10}$~mbar)} & \multicolumn{3}{m{.27\linewidth}|}{ \textit{ex-situ} \newline (after 6 hours \newline in ambient air)}\\ \hline
Layer& \centering $t$ (\AA) & \centering $\sigma$ (\AA)& \centering $t$ (\AA) & \centering $\sigma$ (\AA)& \\ \hline
 CoO$_x$  & --&--&9&4& \\ \hline
Co  & 30&6&26&6&\\ \hline
 Pt  & 205&7&205&7&\\ \hline
\end{tabular}
\end{center}
\caption{Thickness ($t$) and roughness ($\sigma$) of the CO$_x$, Co (sum of Co$_1$,Co$_2$ and Co$_3$ layer) and Pt layers of the same sample measured using PNR measurements in UHV and ex-situ, respectively. The error in each value is $\pm 1$~\AA.}
\label{tab:data} 
\end{table}
For further insight, the UHV and \textit{ex-situ} PNR data  were fitted, for the resulting parameters see Table~\ref{tab:data}. For all fitting parameters, the error was found to be $\pm 1$ \AA. For the \textit{ex-situ} measured PNR data  an additional Co oxide layer (CoO$_x$) on top of the Co layer had to be taken into account. It is well known that Co quickly oxidizes in air.

Figs.~\ref{fig:PNR-CoO}(c) and (d) show the depth profiles of the effective nuclear SLD, effective magnetic SLD and effective magnetic moment per Co atom obtained from the fitting. 
 In order to gain higher accurracy in the magnetic profile, the Co film was subdivided into three layers (Co$_1$, Co$_2$ and Co$_3$) of 10~\AA\ thickness. We allowed the nuclear and magnetic SLD of layer Co$_1$ (in contact with vacuum or air) and layer Co$_3$  (in contact with the Pt film) to vary from its nominal value while for layer Co$_2$  (middle part of Co film) the SLDs were fixed to their nominal values. 

In total,  we may conclude that the UHV transfer chamber is fully working as the transferred Co thin film is 
 not oxidized during its measurement at MARIA.

\section{Summary and conclusion}
\label{sec:conclusion}

In this paper we present the setup and applicability of a versatile 
transfer chamber for measuring  polarized neutron
reflectometry under UHV conditions. The setup is described in detail, being assembled by
standard DN 40 CF UHV parts.  Only one transfer is necessary from
sample fabrication to PNR measurement which may be performed back and
forth. Hence, also PNR studies of the same thin film with increasing
thickness or of multilayer samples are possible.

The  polarized neutron reflectometry  measurements were performed  at the neutron reflectometer MARIA on  30 \AA\   Co thin film grown on 205 \AA\ Pt/ MgO(001)  of $1 \times 1$~cm$^2$ and transferred  to MARIA under UHV conditions. The evaluation of the data reveals no oxidation of the Co film, showing the stability of this chamber for PNR measurements under UHV conditions. 

The general setup of the transfer chamber is versatile and may easily be adopted to other large scale instruments. 

\begin{acknowledgments}
Technical help of S.\ Staringer is gratefully acknowledged.
This project is part of the nanoscience foundry and fine analysis project (NFFA)\cite{NFFA} and  has received funding from the EU's H2020 research and innovation programme under grant agreement n.\ 654360.
\end{acknowledgments}

\bibliography{lit-transfer}

\end{document}